%% file: main_en_Macromolecules_proofreading.tex
\documentclass[a4paper,11pt]{article}

\usepackage[T1]{fontenc}
\usepackage{lmodern}
\usepackage[american]{babel}
\usepackage{graphicx}
\usepackage{amsmath,amssymb}
\usepackage{booktabs}
\usepackage{bm}
\usepackage{url}
\usepackage{cite}
\usepackage{lineno}
\usepackage{float}
\usepackage{color}
\usepackage{caption}
\newcommand{\argmax}{\mathop{\mathrm{arg\,max}}\limits}

\title{Resolving the Effects of the Mixing Process on Hierarchical Structures in Polymer Nanocomposites by Bayesian Ultra-Small-Angle X-ray Scattering Analysis}
\author{
  \parbox{0.95\textwidth}{\centering
  Yui Hayashi$^{1}$, Kazuki Mita$^{2}$, Shigeo Kuwamoto$^{3}$, and Masato Okada$^{1,*}$\\[0.5em]
  \small $^{1}$Graduate School of Frontier Sciences, The University of Tokyo, Kashiwa, Chiba 277-8561, Japan\\
  \small $^{2}$Neutron Science and Technology Center, Comprehensive Research Organization for Science and Society (CROSS), Tokai, Ibaraki 319-1106, Japan\\
  \small $^{3}$Japan Synchrotron Radiation Research Institute (JASRI), Sayo, Hyogo 679-5198, Japan\\[0.5em]
  \small $^{*}$Corresponding Author: okada@edu.k.u-tokyo.ac.jp
  }
}
\date{}

\begin{document}

\maketitle

\begin{abstract}
Bayesian inference was combined with Unified Fit to resolve the effects of the mixing history on polymer nanocomposites whose ultra-small-angle X-ray scattering profiles are nearly indistinguishable by conventional analysis. 
Bayesian Unified Fit was applied to comparing the aggregate size, internal mass-fractal structure, and surface roughness for three carbon black--filled ethylene propylene diene copolymer rubber samples prepared by different mixing processes. 
Whereas conventional Unified Fit reported essentially identical carbon-black dispersions, the radius of gyration, mass fractal dimension, and surface fractal dimension for the three samples all followed the same ordering, which coincides with the reported ordering of the tensile strength and elongation at break and suggests that the inferred aggregate hierarchy is physically meaningful for the macroscopic mechanical response. 
These results identify the aggregate size as a candidate structural descriptor of rubber reinforcement and provide an uncertainty-aware basis for examining processing--structure--property relationships in polymer nanocomposites.
\end{abstract}

\section{Introduction}

Carbon black (CB)--filled ethylene propylene diene copolymer (EPDM) rubber is a representative polymer nanocomposite whose macroscopic performance depends not only on the filler loading but also on the hierarchical organization of the filler within the polymer matrix. 
CB aggregates and networks typically span nanometer to sub-micrometer length scales, and their size, internal packing, surface structure, and connectivity are affected by the mixing history. 
Because these structural features are closely related to the reinforcement of filled rubbers and other polymer nanocomposites, the quantitative characterization of processing-induced filler morphology is a central issue in polymer materials design~\cite{Ikeda2009, Rishi2018, Takenaka2013}.
Small-angle scattering provides a nondestructive route to probing such hierarchical structures over a wide range of length scales~\cite{GuinierFournet1955, GlatterKratky1982, FeiginSvergun1987, Takenaka2013, Takenaka2021, Mashita2013}.

Beaucage and Schaefer introduced the unified Guinier/power-law approach to extract fractal and hierarchical structural information from small-angle scattering profiles, which was subsequently formulated by Beaucage as the unified exponential/power-law approach (hereafter referred to as Unified Fit)~\cite{BeaucageSchaefer1994, Beaucage1995, Beaucage1996, Koga2008}.
By combining Guinier and power-law terms, Unified Fit estimates structural parameters such as the radius of gyration $R_g$, mass fractal dimension $D_m$, and surface fractal dimension $D_s$~\cite{Beaucage1995, Beaucage1996} and has been incorporated in common small-angle scattering analysis packages such as Irena~\cite{IlavskyJemian2009} and SasView~\cite{SasView2025}.
For polymer nanocomposites whose scattering profiles are visually similar, however, conventional point-estimate fitting has difficulty determining whether small differences in these structural parameters are statistically meaningful. 
Gradient-based fitting methods such as quasi-Newton algorithms can also be sensitive to initial values and local optima, which limits their reproducibility when parameters are strongly correlated~\cite{Hammouda2010}.

Bayesian inference provides a natural framework for addressing these limitations because it represents structural parameters as posterior distributions rather than only as point estimates.
Bayesian approaches have been applied to small-angle scattering through maximum-entropy particle-size estimation as a Bayesian regularization of indirect Fourier transformation~\cite{Potton1988}, hyperparameter estimation~\cite{Hansen2000}, probabilistic treatment of structural quantities such as maximum particle size~\cite{Vestergaard2006}, and quantitative assessment of experimental noise~\cite{Larsen2021}.
For scattering models, prior information can suppress overfitting and stabilize estimation under low signal-to-noise conditions~\cite{Larsen2018}, and global fitting can provide realistic parameter correlations and error analyses over the measured $q$ range~\cite{Frewein2019}.
In a Bayesian inference framework, a data-generation model that incorporates measurement noise can be combined with physically motivated prior distributions, and Bayes' theorem will yield posterior distributions of the parameters~\cite{Hayashi2023, Hayashi2024, Oyama2025}.
Replica exchange Monte Carlo (REMC) sampling~\cite{HukushimaNemoto1996, EarlDeem2005} improves exploration of multimodal posteriors while Bayesian model selection based on posterior model probabilities or Bayesian free energy enables an objective comparison of candidate noise models~\cite{Nagata2012, Hayashi2024}.

In this study, we combined Bayesian inference with Unified Fit not only to estimate structural parameters but also to test whether apparently similar ultra-small-angle X-ray scattering (USAXS) profiles of EPDM/CB samples retain statistically resolvable signatures of the mixing history. 
We analyzed three vulcanized EPDM/CB samples prepared by different mixing processes and previously reported by Nakanishi \textit{et al.}~\cite{Nakanishi2021} for which conventional Unified Fit analysis indicated that the CB dispersion states were essentially identical.
Here, we compared posterior distributions of $R_g$, $D_m$, and $D_s$ to quantify processing-induced differences in the CB aggregate size, internal packing, and surface roughness.
Because preprocessing may alter the noise statistics of small-angle scattering data, we evaluated six candidate noise models and selected the most appropriate by Bayesian inference before comparing the structural posteriors.
The resulting structural trends were then compared with the reported tensile properties to clarify the extent to which the inferred hierarchical filler structures are physically meaningful for rubber reinforcement.

The remainder of this paper is organized as follows.
Section~\ref{sec:bayes} introduces the Bayesian inference framework for parameter estimation and model selection.
Section~\ref{sec:model} formulates the Unified Fit model and the physical meaning of its parameters.
Section~\ref{sec:experiment} describes the candidate noise models, the EPDM/CB samples, the experimental data, and the validation protocol.
Section~\ref{sec:results} presents the results and discussion.
Section~\ref{sec:conclusion} concludes the paper and outlines future work.

\section{Formulation of the Bayesian Inference Framework}
\label{sec:bayes}

We treated the small-angle scattering profiles within a Bayesian inference framework for (i) noise-model selection and (ii) parameter estimation~\cite{Nagata2012, Hayashi2023, Hayashi2024, Oyama2025}.
Measured data are denoted by $\mathcal{D}=\{q_i, y_i\}_{i=1}^N$, where $q_i$ is the scattering vector, $y_i$ is the scattering intensity and $N$ is the number of data points.

\subsection{Bayesian Inference}
\label{subsec:bayes_inference}

$K$ is an index characterizing the data-generation model, and $\Theta$ represents the corresponding parameters.
The data-generation model is a probabilistic model of the measurement noise, which can be referred to as the noise model $K_{\mathrm{noise}}$.
Let $f(q,\Theta)$ denote the scattering intensity model; then, the generative probability of the observation $y_i$ at $q_i$ for the data-generation model $K$ is $p(y_i|q_i,\Theta,K)$.
We define the functional form of this generative probability as the noise model.
The vector $\Theta$ comprises both the structural parameters and, where appropriate, the noise parameters.

Assuming that the \(N\) observations are independent, the likelihood factorizes as
\begin{equation}
    p(\mathcal{D}|\Theta, K) = \prod^N_{i=1} p(y_i|q_i,\Theta, K). \label{likelihood}
\end{equation}
We define the cost function $E(\Theta,K)$ as
\begin{equation}
    E(\Theta,K) = -\frac{1}{N}\log p(\mathcal{D}|\Theta,K). \label{def_cost}
\end{equation}
Let $\varphi(K)$ be the prior over the model index $K$ and $\varphi(\Theta|K)$ be the prior over the parameters $\Theta$ given $K$.
By Bayes' theorem, the posterior distribution of the parameters given the observed data $\mathcal{D}$ is
\begin{equation}
    p(\Theta|\mathcal{D}, K) = \frac{\exp\left(-N E(\Theta, K)\right)\varphi(\Theta|K)}{Z(K)},
\end{equation}
where $Z(K)$ is the marginal likelihood serving as the normalizing constant of the posterior:
\begin{equation}
    Z(K) = \int \exp\left(-N E(\Theta, K)\right)\varphi(\Theta|K)\, d\Theta. \label{target_marginal_likelihood}
\end{equation}

The posterior probability of model $K$ is given by
\begin{eqnarray}
    p(K|\mathcal{D}) &=& \frac{\exp(-F(K))\varphi(K)}{\sum_{k} \exp(-F(k))\varphi(k)}, \label{model_likelihood}\\
    F(K) &=& -\log Z(K), \label{free_energy}
\end{eqnarray}
where $F(K)$ is the Bayesian free energy~\cite{Watanabe2009} and reflects the trade-off between goodness of fit and model complexity.
Computing $p(K|\mathcal{D})$ for each candidate model $\{K\}$ and comparing the resulting probabilities therefore provides a quantitative criterion for model selection.

The marginal likelihood $Z(K)$ generally involves a multidimensional integral and cannot be evaluated analytically.
Here we evaluate $Z(K)$ numerically by using the replica exchange Monte Carlo (REMC) method~\cite{Nagata2012, HukushimaNemoto1996}.
We introduce $L$ inverse temperatures $\beta_1, \ldots, \beta_L$ that satisfy
\begin{equation}
    0=\beta_1 < \cdots < \beta_L = 1. \label{beta_relation}
\end{equation}
Then, we use a Markov chain Monte Carlo (MCMC) algorithm to draw samples at each $\beta_l$ ($1 \leq l \leq L$) from the distribution
\begin{equation}
    p(\Theta_l|\mathcal{D}, K, \beta_l) \propto \exp\left(-N\beta_l E(\Theta_l, K)\right)\varphi(\Theta_l |K). \label{posterior_dist}
\end{equation}
Replicas at adjacent inverse temperatures are exchanged with the probability that satisfies detailed balance:
\begin{equation}
    p(\Theta_l \leftrightarrow \Theta_{l+1}) = \min\!\left[1,\ \exp\bigl(N(\beta_{l+1} - \beta_{l})(E(\Theta_{l+1}, K) - E(\Theta_{l}, K))\bigr) \right].
\end{equation}
This exchange mechanism enables global sampling and helps avoid entrapment in local optima.

Let $Z(K, \beta) = \int \exp(-N \beta E(\Theta, K))\varphi(\Theta|K)\, d\Theta$ denote the marginal likelihood at the inverse temperature $\beta$. 
Then, the target marginal likelihood $Z(K) = Z(K, \beta=1)$ can be expressed as
\begin{equation}
    Z(K, \beta=1) = \prod_{l=1}^{L-1} \left\langle \exp\!\left(-N(\beta_{l+1} - \beta_{l}) E(\Theta, K) \right) \right\rangle_{p(\Theta_l|\mathcal{D}, K, \beta_l)}, \label{final_marginal_likelihood}
\end{equation}
where $\langle \cdot \rangle_{p(\Theta_l|\mathcal{D}, K, \beta_l)}$ denotes the expectation with respect to the distribution $p(\Theta_l|\mathcal{D}, K, \beta_l)$~\cite{Nagata2012}.
Equation~(\ref{final_marginal_likelihood}) presupposes that the prior is normalized ($\int \varphi(\Theta|K)\, d\Theta = 1$, i.e.\ $Z(K, \beta=0) = 1$).
In this study we used bounded uniform priors normalized by their support volume.
By evaluating Equation~(\ref{final_marginal_likelihood}) on samples obtained by REMC, we can estimate $Z(K)$, and from this  we can obtain the Bayesian free energy $F(K)$ and the model posterior $p(K|\mathcal{D})$.
The  Metropolis algorithm is used as the MCMC sampler at each inverse temperature~\cite{Metropolis1953}.

\subsection{Estimation of Model Parameters}

The marginal-likelihood computation outlined by Equations~(\ref{beta_relation})--(\ref{final_marginal_likelihood}) also yields the parameter posterior at $\beta=1$, $p(\Theta_L|\mathcal{D}, K, \beta_L=1)$.
This distribution is precisely the Bayesian estimate of the parameters~\cite{Hayashi2023}.
In other words, Bayesian model selection of noise model candidates naturally estimates the parameters as a byproduct.
Because the posterior is sampled with REMC, this procedure facilitates global exploration of the posterior and reduces sensitivity to local optima.
Summary statistics and the shape of the sampled posterior also provide direct measures of the reliability of the estimates.
A point estimate is given by the maximum a posteriori (MAP) solution.
The MAP estimate of the parameters of model $K$ maximizes the posterior at $\beta=1$:
\begin{eqnarray}
    \Theta_\text{MAP} &=& \argmax_{\Theta}\ p(\Theta|\mathcal{D},K), \\
    &=& \argmax_{\Theta}\ \exp\!\left( -N E(\Theta, K) \right) \varphi(\Theta|K). \label{MAP_solution}
\end{eqnarray}

\section{Unified Fit Model and Structural Parameters}
\label{sec:model}

Unified Fit uses the Beaucage unified scattering function or unified exponential/power-law approach~\cite{Beaucage1995, Beaucage1996}.
The magnitude of the scattering vector is $q = 4\pi \sin\theta/\lambda_\mathrm{w}$, where $\lambda_\mathrm{w}$ is the incident wavelength and $2\theta$ is the scattering angle.
We employed the three-term scattering intensity model used by Nakanishi \textit{et al.}~\cite{Nakanishi2021} to analyze the CB dispersions in their EPDM/CB samples:
\begin{eqnarray}
    f(q) &=& A\, q^{-D_m} \exp\!\left(-\frac{q^2 R_g^2}{3}\right)
    + B \exp\!\left(-\frac{q^2 R_g^2}{3}\right) \nonumber\\
    &&+ C\, q^{-(2d-D_s)} \left[\mathrm{erf}\!\left(\frac{q R_g}{\sqrt{6}}\right)\right]^{3(2d-D_s)},
    \label{eq:unified_fit}
\end{eqnarray}
where $d$ denotes the spatial dimension (set to $d=3$).
The model contains six structural parameters: $A$ is the amplitude of the power-law term, $B$ is the amplitude of the Guinier term, $C$ is the amplitude of the high-$q$ scattering term, $D_m$ is the mass fractal dimension, $D_s$ is the surface fractal dimension, and $R_g$ is the radius of gyration.
The three terms in Equation~(\ref{eq:unified_fit}) correspond to distinct structural features.
The first term represents mass-fractal scattering with a Guinier cutoff and describes the power-law decay in the low-$q$ region.
The second term is a Guinier (exponential) decay carrying information about the overall size of the scatterer.
The third term describes the surface scattering in the high-$q$ region,  with the error function ($\mathrm{erf}$) smoothly bridging the Guinier and power-law regimes~\cite{Beaucage1995}.
For the EPDM/CB samples, $R_g$ was interpreted as a representative size of CB aggregates, $D_m$ was interpreted as a measure of the mass distribution or internal packing within the aggregates, and $D_s$ was interpreted as a measure of surface roughness or interfacial structure.
We can collect the structural parameters in the vector
\begin{equation}
    \Theta_{\mathrm{struct}} = \{A, B, C, D_m, D_s, R_g\}.
\end{equation}
If we denote the noise parameters as $\Theta_{\mathrm{noise}}$, then the full parameter vector introduced in Section~\ref{sec:bayes} becomes $\Theta = \{\Theta_{\mathrm{struct}},\, \Theta_{\mathrm{noise}}\}$.

\section{Datasets and Analysis Setup}
\label{sec:experiment}

\subsection{Candidate Noise Models}
\label{sec:noise}

In general, the data-generation process is described by a probabilistic model that incorporates measurement noise.
Because the small-angle scattering intensity is recorded by a detector as photon or neutron counts, the raw data are theoretically governed by Poisson statistics~\cite{GlatterKratky1982, Nagata2019, Katakami2022}.
However, preprocessing steps such as transmission and background corrections are often applied, and the resulting data are typically treated as continuous-valued.
Therefore, the statistical character of the raw counts is not necessarily preserved.
We accordingly considered six candidate noise models and selected the most appropriate for the data through Bayesian model selection.
For each model, $f_i \equiv f(q_i, \Theta)$ was used to denote the predicted scattering intensity.

\vspace{1em}
\noindent\textbf{(a) Gaussian (constant variance)}: The variance of the scattering intensity is assumed to be constant for all data points:
\begin{equation}
    p\left(y_i \mid q_i,\Theta,K\right)
    = \mathcal{N}\!\left(y_i \mid f_i,\ \sigma_0^2\right). \label{gaussian_noise}
\end{equation}
The corresponding cost function is
\begin{equation}
    E(\Theta,K)
    = \frac{1}{2N}\sum_{i=1}^{N}\left\{\frac{(y_i-f_i)^2}{\sigma_0^2} + \log(2\pi\sigma_0^2)\right\}, \label{gaussian_cost}
\end{equation}
and $\sigma_0$ is added to the parameter set.

\vspace{1em}
\noindent\textbf{(b) Poisson--Gaussian (Poisson approximation)}: A Gaussian approximation in which the variance is set equal to the mean to reflect the equality of mean and variance for the Poisson distribution:
\begin{equation}
    p\left(y_i \mid q_i,\Theta,K\right)
    = \mathcal{N}\!\left(y_i \mid f_i,\ f_i\right). \label{poisson_gaussian_noise}
\end{equation}
This yields
\begin{equation}
    E(\Theta,K)
    = \frac{1}{2N}\sum_{i=1}^{N}\left\{\frac{(y_i-f_i)^2}{f_i} + \log(2\pi f_i)\right\}. \label{poisson_gaussian_cost}
\end{equation}
This model has no free noise parameters, so it is reduced to only the structural parameters.
The Gaussian approximation is appropriate when $f_i$ is sufficiently large. In the high-$q$ regions, where the intensity is low, its accuracy may degrade.

\vspace{1em}
\noindent\textbf{(c) Scaled Poisson}: The variance is proportional to the predicted intensity, with the proportionality coefficient $\alpha_1 > 0$:
\begin{eqnarray}
    p\left(y_i \mid q_i,\Theta,K\right)
    &=& \mathcal{N}\!\left(y_i \mid f_i,\ \sigma^2(q_i)\right), \label{scaled_poisson_noise}\\
    \sigma^2(q_i) &=& \alpha_1 f_i.
\end{eqnarray}
Although $\sigma^2(q_i)$ also depends on the model parameters $\Theta$, we suppress this dependence in the notation for brevity (as well as for the following noise models). 

\vspace{1em}
\noindent\textbf{(d) Poisson + correction}: A correction term proportional to the squared predicted intensity is added to the Poisson term~\cite{Sakishita2025}:
\begin{equation}
    \sigma^2(q_i) = f_i + \alpha_1 f_i^2. \label{poisson_correction_noise}
\end{equation}

\vspace{1em}
\noindent\textbf{(e) Corrected Gaussian 2}: This is a two-parameter model with a constant term and a term proportional to the squared predicted intensity:
\begin{equation}
    \sigma^2(q_i) = \alpha_1 + \alpha_2 f_i^2. \label{corrected_gaussian2_noise}
\end{equation}

\vspace{1em}
\noindent\textbf{(f) Corrected Gaussian 3}: This is a three-parameter model that includes a constant term, a term proportional to the predicted intensity, and a term proportional to the squared predicted intensity~\cite{Nagai2020}:
\begin{equation}
    \sigma^2(q_i) = \alpha_1 + \alpha_2 f_i + \alpha_3 f_i^2. \label{corrected_gaussian3_noise}
\end{equation}

\vspace{0.5em}
\noindent Models (c)--(f) all assume a Gaussian likelihood $p(y_i \mid q_i,\Theta,K) = \mathcal{N}(y_i \mid f_i,\, \sigma^2(q_i))$ and differ only in the definition of $\sigma^2(q_i)$.
Therefore, their cost functions share a common form:
\begin{equation}
    E(\Theta,K)
    = \frac{1}{2N}\sum_{i=1}^{N}\left\{\frac{(y_i-f_i)^2}{\sigma^2(q_i)} + \log(2\pi\sigma^2(q_i))\right\}. \label{general_cost}
\end{equation}

We set $K_{\mathrm{noise}}\in\{(\mathrm{a}), (\mathrm{b}), (\mathrm{c}), (\mathrm{d}), (\mathrm{e}), (\mathrm{f})\}$, evaluated $p(K|\mathcal{D})$ from Equation~(\ref{model_likelihood}), and selected the candidate noise model that maximized the posterior probability.

\subsection{Samples and Measurement}

We analyzed the USAXS profiles of three vulcanized EPDM/CB samples reported by Nakanishi \textit{et al.}~\cite{Nakanishi2021}.
Samples S1--S3 were prepared by different mixing processes:

\begin{itemize}
    \item \textbf{S1:} Banbury-type mixer (tangential rotors) with a mixing time of 5\,min (Process~I).
    \item \textbf{S2:} Kneader (intermeshing rotors) with a mixing time of 5\,min (Process~II).
    \item \textbf{S3:} 8-inch two-roll mill with a mixing time of 30\,min (Process~III).
\end{itemize}

\noindent The base polymer was EPDM (4045M; Mitsui Chemicals Inc.) compounded with carbon black (CB; HAF Asahi \#70), paraffin process oil, zinc oxide (ZnO), stearic acid, vulcanization accelerators (2-mercaptobenzothiazole and tetramethylthiuram disulfide), and sulfur~\cite{Nakanishi2021}.

USAXS measurements were performed at beamlines BL19B2 and BL03XU at SPring-8 (Harima, Hyogo, Japan)~\cite{Nakanishi2021}.
For each sample, the scattering profile comprised $N=814$ data points expressed as $(q, I(q))$ pairs.
Full details of the sample preparation, compounding ratios, and experimental setups for the USAXS measurements are described by Nakanishi \textit{et al.}~\cite{Nakanishi2021}.

\subsection{Prior Distribution Settings}

A uniform prior was used for every parameter.
The prior ranges for each candidate noise model are listed in Table~\ref{tab:prior_all}.
The ranges were chosen to span physically reasonable values.

\begin{table}[H]
    \centering
    \caption{Prior distributions of the parameters for each candidate noise model. All priors are uniform $\mathrm{U}[\text{lower},\text{upper}]$.}
    \label{tab:prior_all}
    \footnotesize
    \setlength{\tabcolsep}{4pt}
    \begin{tabular}{lcccccc}
      \toprule
      Parameter & (a) & (b) & (c) & (d) & (e) & (f) \\
      \midrule
      $A$    & $[1, 3]$ & $[1, 3]$ & $[1, 3]$ & $[1, 3]$ & $[1, 3]$ & $[1, 3]$ \\
      $B$    & $[2800, 3200]$ & $[600, 2000]$ & $[600, 1000]$ & $[300, 2000]$ & $[300, 2000]$ & $[300, 2000]$ \\
      $C$    & $[10^{-4}, 10]$ & $[10^{-4}, 1]$ & $[10^{-4}, 1]$ & $[10^{-4}, 1]$ & $[10^{-4}, 1]$ & $[10^{-4}, 1]$ \\
      $D_m$  & $[1, 3]$ & $[1, 3]$ & $[1, 3]$ & $[1, 3]$ & $[1, 3]$ & $[1, 3]$ \\
      $D_s$  & $[2, 3]$ & $[2, 3]$ & $[2, 3]$ & $[2, 3]$ & $[2, 3]$ & $[2, 3]$ \\
      $R_g$ [nm] & $[20, 50]$ & $[20, 40]$ & $[20, 40]$ & $[20, 40]$ & $[20, 40]$ & $[20, 40]$ \\
      \midrule
      $\sigma_0$ & $[500, 600]$ & --- & --- & --- & --- & --- \\
      $\alpha_1$ & --- & --- & $[0, 10]$ & $[0, 0.1]$ & $[0, 0.1]$ & $[0, 0.1]$ \\
      $\alpha_2$ & --- & --- & --- & --- & $[0, 0.1]$ & $[0, 0.1]$ \\
      $\alpha_3$ & --- & --- & --- & --- & --- & $[0, 0.1]$ \\
      \bottomrule
    \end{tabular}
\end{table}

\subsection{Synthetic Data for Validation}
\label{subsec:synthetic}

To validate the proposed method, we performed estimation experiments on synthetic data.
The point estimates obtained for the S1 USAXS profile by the conventional non-linear least-squares method were taken as the true structural parameters $\Theta_{\mathrm{true}}$, and the corresponding true scattering intensity $f_{\mathrm{true}}(q_i) = f(q_i, \Theta_{\mathrm{true}})$ was generated from Equation~(\ref{eq:unified_fit}).
The true values were set to $A=2.097$, $B=702.4$, $C=0.071$, $D_m=2.226$, $D_s=2.718$, and $R_g=24.43$\,nm.
Noise was added according to candidate noise model (e) (Equation~(\ref{corrected_gaussian2_noise})).
A scalar coefficient $\lambda > 0$ was used to control the noise intensity through
\begin{equation}
    \sigma^2(q_i) = \lambda \left(\alpha_{1,0} + \alpha_{2,0}\, f_{\mathrm{true}}(q_i)^2 \right),
    \label{eq:synth_noise}
\end{equation}
where $\alpha_{1,0} = 3.0 \times 10^{-4}$ and $\alpha_{2,0} = 4.0 \times 10^{-4}$ were the baseline noise parameters. 
A larger $\lambda$ corresponds to a lower signal-to-noise ratio.
We considered two noise levels: $\lambda = 10$ and $\lambda = 100$.
Synthetic observations were generated as $y_i = f_{\mathrm{true}}(q_i) + \varepsilon_i$ with $\varepsilon_i \sim \mathcal{N}(0, \sigma^2(q_i))$.
Estimation was performed by using candidate noise model (e).

\section{Results and Discussion}
\label{sec:results}

This section first validates the reliability of the analysis framework using synthetic data.
We then determine an appropriate noise model for the experimental USAXS data via Bayesian model selection.
On this validated basis, we examine the central question of this study---whether the hierarchical filler structures in EPDM/CB retain statistically resolvable signatures of the mixing process---by comparing structural posterior distributions among the three samples.

\subsection{Validation against Synthetic Data}

\begin{figure}[H]
  \centering
  \includegraphics[width=\textwidth]{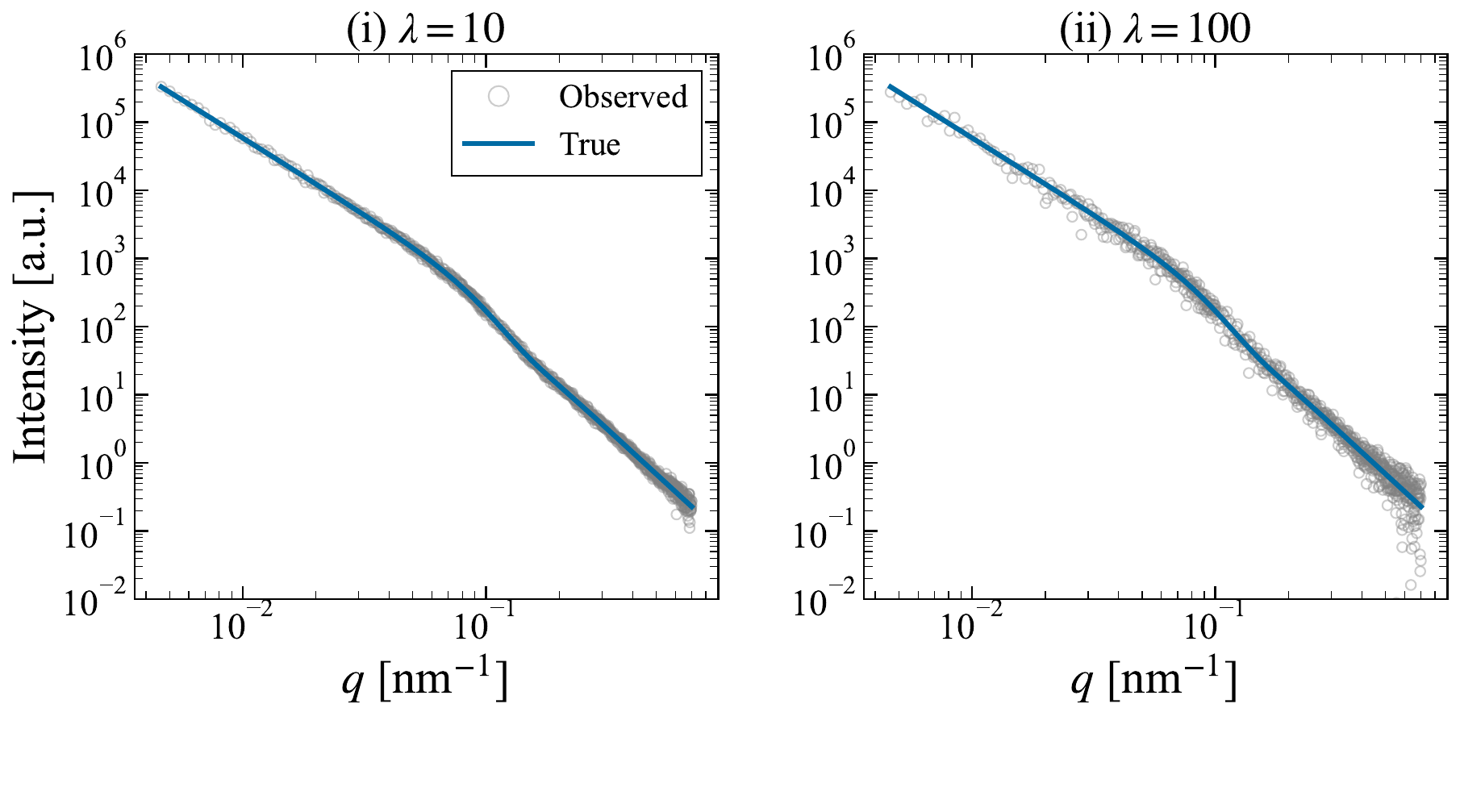}
  \caption{Synthetic data and the underlying true scattering curve. (i) $\lambda=10$ and (ii) $\lambda=100$. The gray circles indicate noisy synthetic data, and the blue solid line indicates the true scattering curve.}
  \label{fig:synthetic_data}
\end{figure}

To validate the reliability of Bayesian Unified Fit, we performed estimation experiments on synthetic data.
Following the conditions described in Section~\ref{subsec:synthetic}, we analyzed synthetic data generated at two noise levels: $\lambda = 10$ and $\lambda = 100$.
Figure~\ref{fig:synthetic_data} shows the generated synthetic data.
For $\lambda = 10$, the noise was low relative to the scattering intensity and deviation from the true scattering curve was limited.
For $\lambda = 100$, the noise was substantially greater and the deviation from the true curve was more pronounced.

\begin{figure}[H]
  \centering
  \includegraphics[width=\textwidth]{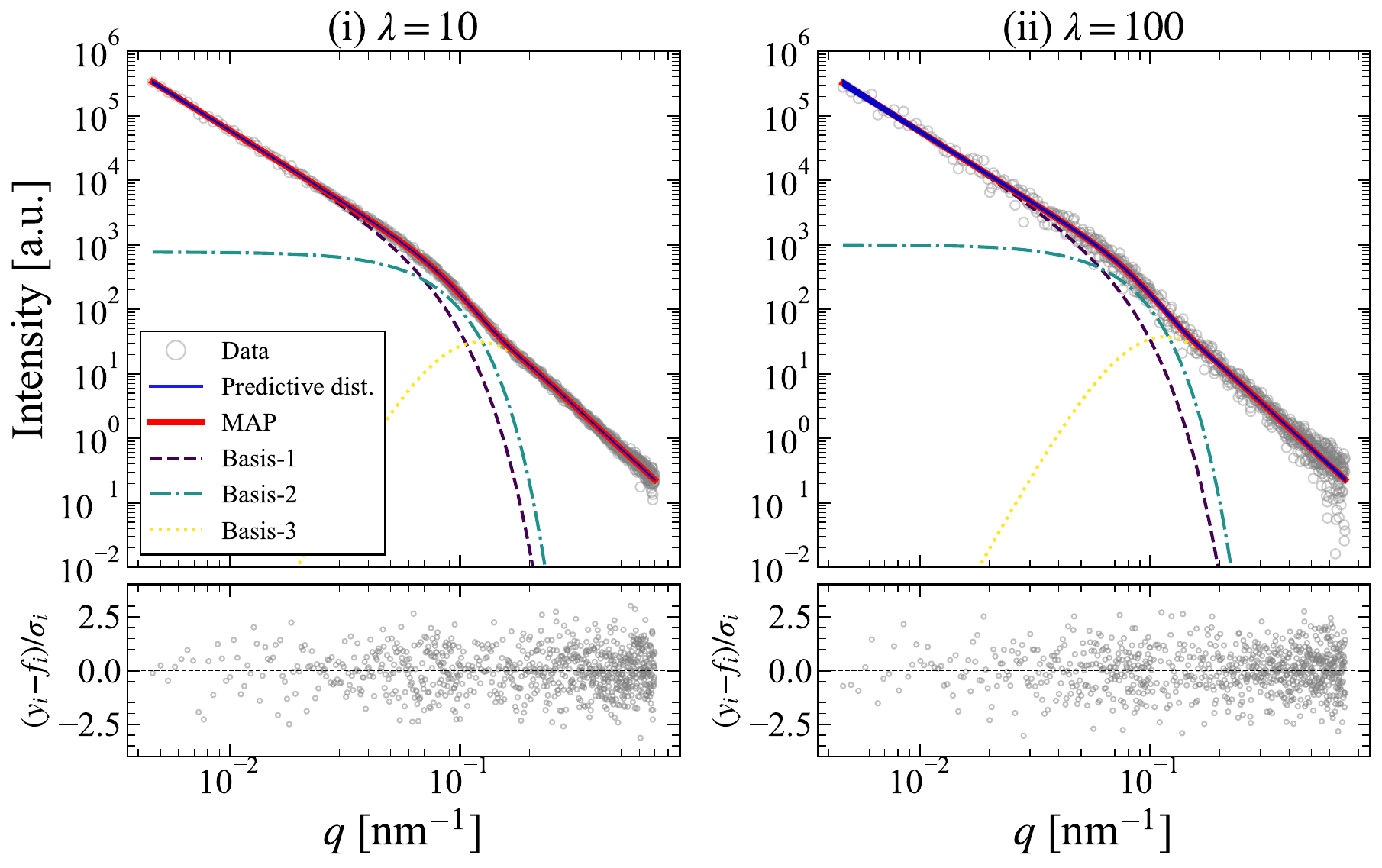}
  \caption{Fitting results for the synthetic data. (i) $\lambda=10$ and (ii) $\lambda=100$. The red solid line indicates the MAP estimate, and the thin blue lines indicate the predictive distribution from 2000 posterior samples. The dashed, dot-dashed, and dotted curves show the first (Basis-1: mass-fractal scattering), second (Basis-2: Guinier scattering), and third (Basis-3: surface scattering) terms, respectively, of Equation~(\ref{eq:unified_fit}) evaluated at the MAP. The lower panels show the standardized residuals.}
  \label{fig:synthetic_fitting}
\end{figure}

\begin{figure}[H]
  \centering
  \includegraphics[width=\textwidth]{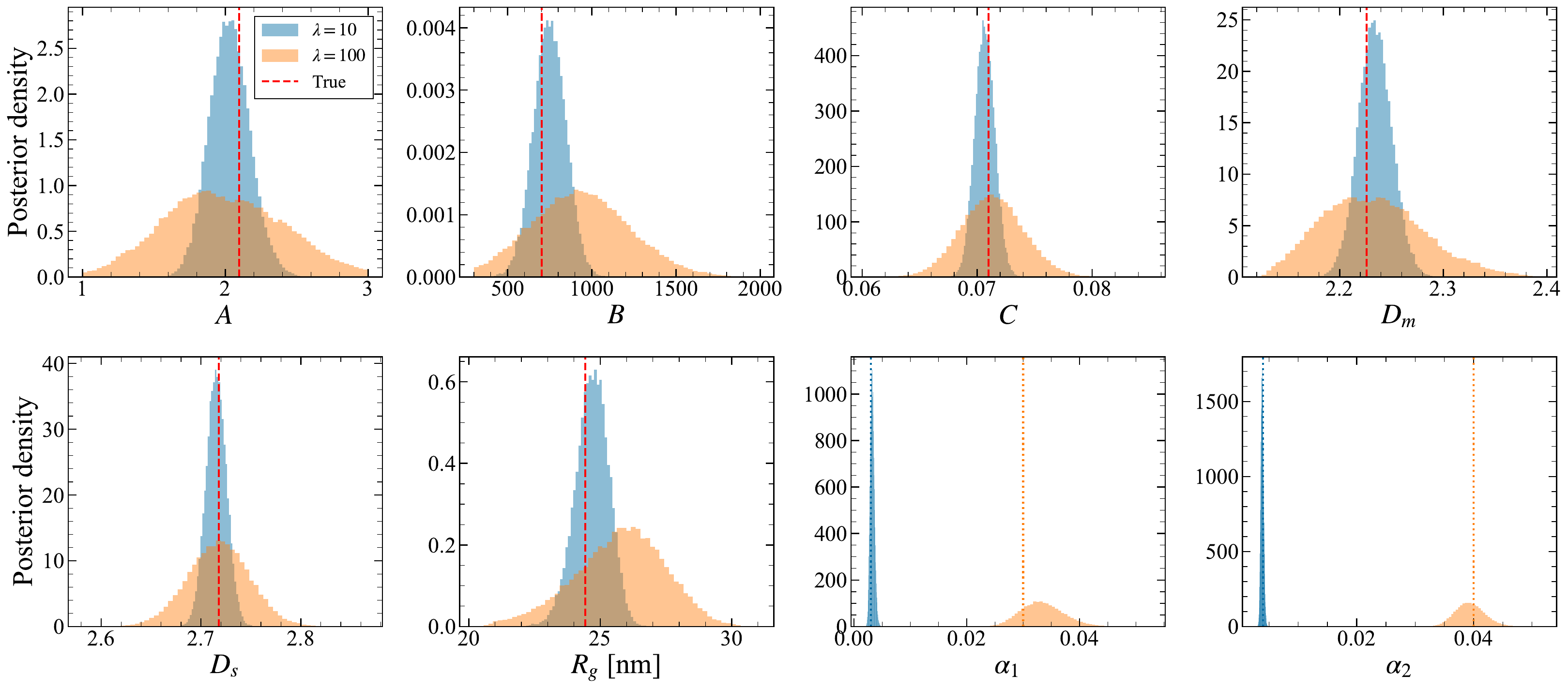}
  \caption{Posterior distributions of the parameters for the synthetic data: $\lambda=10$ (blue) and $\lambda=100$ (orange). Red dashed lines indicate the true values of the structural parameters. Dotted lines indicate the true values of the noise parameters at each $\lambda$.}
  \label{fig:synthetic_posterior}
\end{figure}

Figure~\ref{fig:synthetic_fitting} shows the fitting results for the two noise levels.
The red solid line shows the scattering curve at the MAP estimate, and the thin blue lines show the predictive distribution evaluated on 2000 samples drawn from the posterior.
\noindent
Figure~\ref{fig:synthetic_posterior} shows the parameter posteriors.
The posteriors for $\lambda = 10$ (blue) and $\lambda = 100$ (orange) are overlaid; red dashed lines indicate the true structural parameter values.
For $\lambda = 10$, all structural parameter posteriors were concentrated near the true values indicating high estimation accuracy.
For $\lambda = 100$, the posterior widths broadened markedly, which reflects the increased uncertainty caused by the greater noise.
Even at $\lambda = 100$, however, the posterior peaks still lay close to the true values, which indicates that the MAP estimates remained robust against noise.
The posteriors of the noise parameters $\alpha_1$ and $\alpha_2$ were also concentrated near their true values, which confirms that the parameters of candidate noise model (e) were estimated correctly.

Table~\ref{tab:synthetic_map_ci} reports the MAP estimates and 95\% credible intervals together with the true values for both noise levels.
At $\lambda = 10$, the true values fell within the 95\% credible interval for every structural parameter, which confirms the validity of the estimation.
At $\lambda = 100$, the credible intervals for $A$ and $B$ widened substantially, but the true values still lay within the intervals.
The true noise parameters $\alpha_1$ and $\alpha_2$ likewise lay within their credible intervals for both conditions, which indicates that the parameters of candidate noise model (e) were inferred reliably.

\begin{table}[H]
    \centering
    \caption{MAP estimates and 95\% credible intervals for the synthetic data.}
    \label{tab:synthetic_map_ci}
    \resizebox{\textwidth}{!}{%
    \begin{tabular}{lccc}
      \toprule
      Parameter & True value & $\lambda=10$ & $\lambda=100$ \\
      \midrule
      $A$     & 2.097  & $1.978\ [1.762,\ 2.320]$ & $1.859\ [1.237,\ 2.794]$ \\
      $B$     & 702.4  & $770.6\ [558.6,\ 936.5]$ & $1003\ [419,\ 1528]$ \\
      $C$     & 0.071  & $0.0707\ [0.0690,\ 0.0724]$ & $0.0708\ [0.0661,\ 0.0768]$ \\
      $D_m$   & 2.226  & $2.240\ [2.203,\ 2.267]$ & $2.243\ [2.148,\ 2.336]$ \\
      $D_s$   & 2.718  & $2.716\ [2.695,\ 2.735]$ & $2.714\ [2.656,\ 2.778]$ \\
      $R_g$ [nm] & 24.43 & $24.83\ [23.36,\ 25.85]$ & $26.24\ [22.08,\ 28.83]$ \\
      \midrule
      $\alpha_1$ & $3.0\times 10^{-3}$ / $3.0\times 10^{-2}$ & $3.14\times 10^{-3}\ [2.6,\ 4.0]\times 10^{-3}$ & $3.26\times 10^{-2}\ [2.7,\ 4.1]\times 10^{-2}$ \\
      $\alpha_2$ & $4.0\times 10^{-3}$ / $4.0\times 10^{-2}$ & $3.80\times 10^{-3}\ [3.4,\ 4.3]\times 10^{-3}$ & $3.83\times 10^{-2}\ [3.5,\ 4.5]\times 10^{-2}$ \\
      \bottomrule
    \end{tabular}}
    \par\smallskip
    \begin{minipage}{\textwidth}
      \footnotesize\raggedright
      Notes: Values in brackets denote the lower and upper bounds of the 95\% credible interval.
    \end{minipage}
\end{table}

\subsection{Noise Model Selection and Fitting of the Experimental Data}

\begin{figure}[H]
  \centering
  \includegraphics[width=0.75\textwidth]{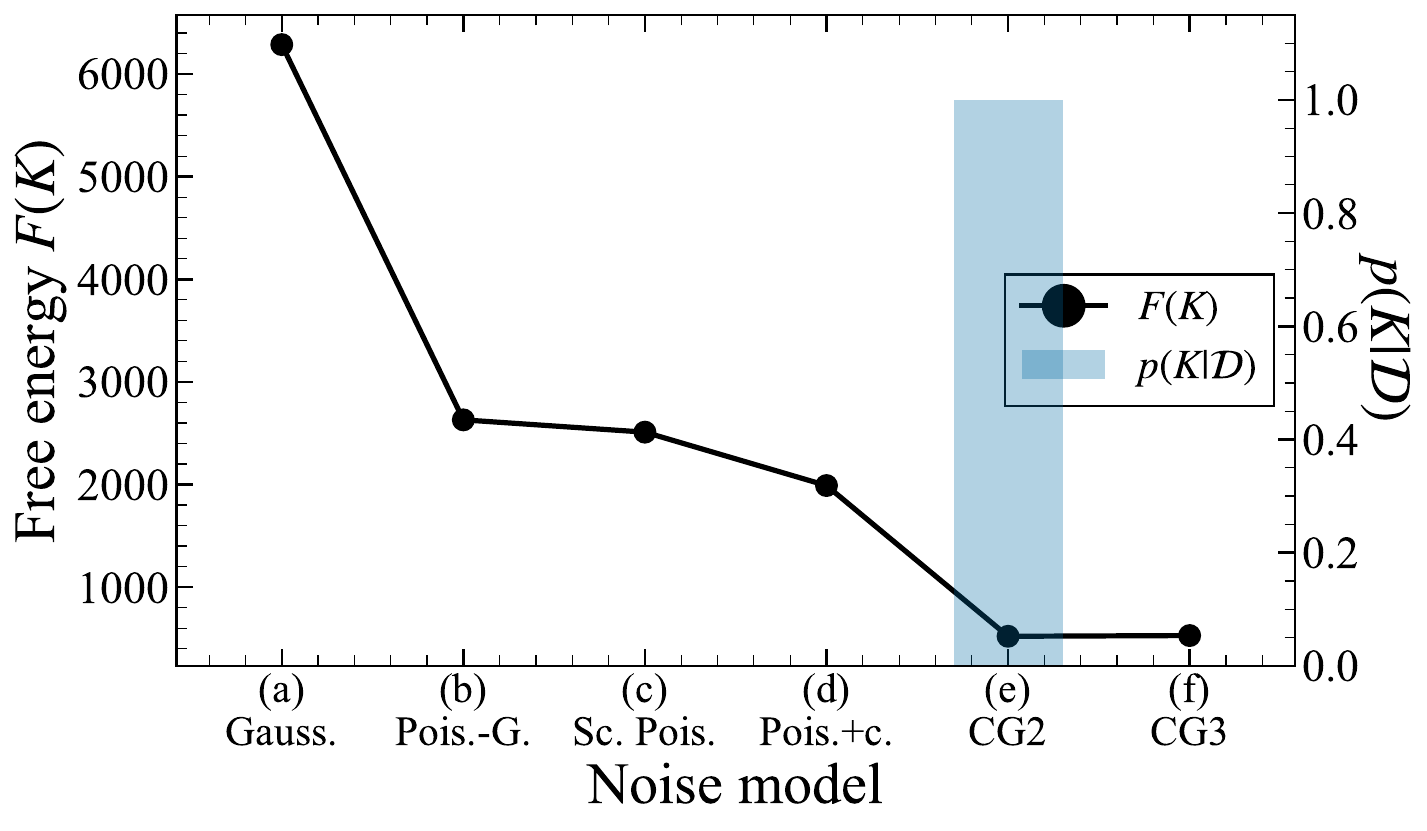}
  \caption{Bayesian free energy $F(K)$ (black circles, solid line; left axis) and posterior probability $p(K|\mathcal{D})$ (blue bars; right axis) of each candidate noise model for the experimental data of sample S1.}
  \label{fig:noise_model}
\end{figure}

Bayesian inference was applied to selecting which of the six candidate noise models fitted the experimental data for sample S1.
Figure~\ref{fig:noise_model} shows the Bayesian free energy $F(K)$ and the posterior probability $p(K|\mathcal{D})$ of each candidate noise model.
The lowest Bayesian free energy was obtained for candidate noise model (e) with $F = 520.97$.
The next-lowest was obtained for candidate noise model (f) with $F = 528.11$.
Although these values may appear close in absolute terms, the difference $\Delta F = 7.14$ corresponds to a Bayes factor of $\exp(7.14) \approx 1.3 \times 10^{3}$ in favor of model (e), so the posterior probability of model (e) is essentially unity under a uniform prior over the models.
Model (f) introduces an additional linear term in the predicted intensity, corresponding to its $\alpha_2 f_i$ contribution, but the data do not support this additional term.
The remaining models (a)--(d) yielded substantially higher $F$ values and were therefore less supported by the experimental data for sample S1.
We should note that these results are specific to the data analyzed in this study, and we do not intend to claim noise model (e) as universally optimal.

\begin{figure}[H]
  \centering
  \includegraphics[width=\textwidth]{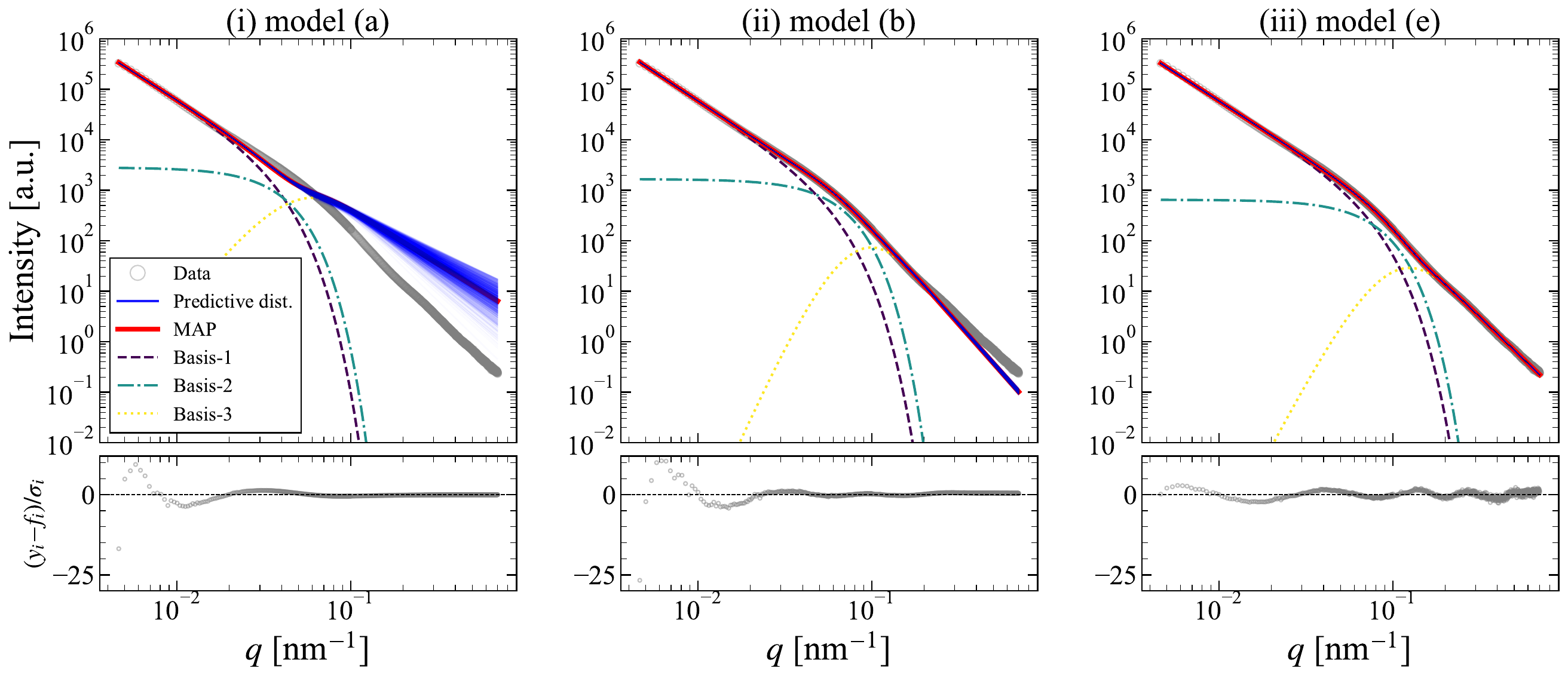}
  \caption{Fitting results on the S1 experimental data: (i) model (a), (ii) model (b), and (iii) model (e). The red solid line indicates the MAP estimate. The thin blue lines indicate the predictive distribution from 2000 posterior samples. Basis-1, Basis-2, and Basis-3 denote the mass-fractal, Guinier, and surface-scattering terms, respectively, of Equation~(\ref{eq:unified_fit}). The lower panels show the standardized residual $(y_i - f_i)/\sigma_i$.}
  \label{fig:real_fitting}
\end{figure}

Figure~\ref{fig:real_fitting} compares the fitting results for the experimental data of sample S1 under three candidate noise models: (i) model (a), (ii) model (b), and (iii) model (e).
In each panel, the red solid line shows the MAP estimate, and the thin blue lines show the predictive distribution from 2000 posterior samples.
For model (a), the assumption of constant variance proved inappropriate. 
The fit was biased toward the high-intensity data points in the low-$q$ region, and the predictive distribution was broad.
For model (b), the situation improved, but the fit remained unsatisfactory in the high-$q$ region.
For model (e), the fitted curve agreed well with the data over the full range of $q$, and the predictive distribution appropriately reflected the observed intensity-dependent scatter.
These results support the use of model (e) for the comparison of structural parameters.

\begin{figure}[H]
  \centering
  \includegraphics[width=\textwidth]{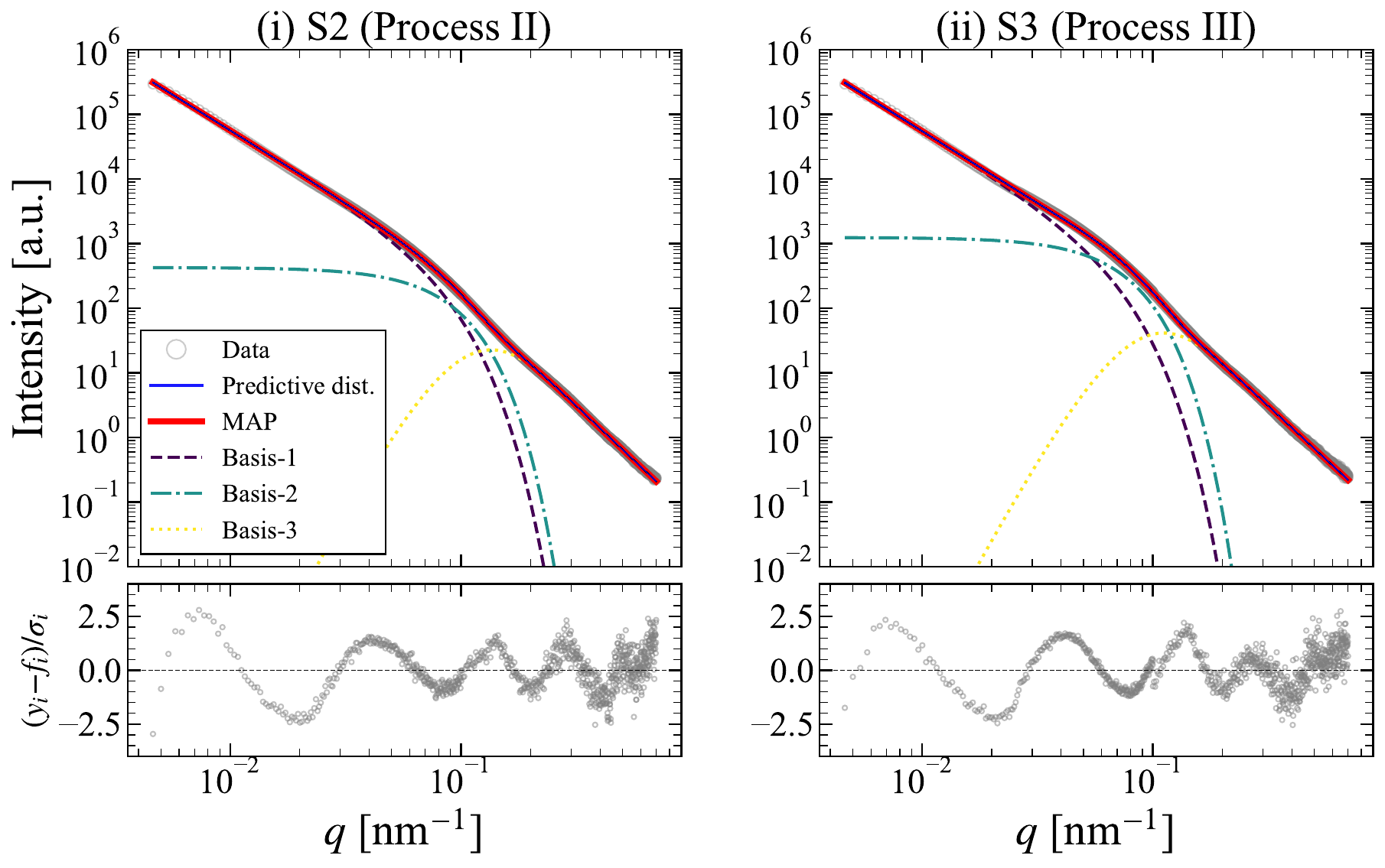}
  \caption{Fitting results to the experimental data of samples (i) S2 and (ii) S3 using candidate noise model (e). The red solid line indicates the MAP estimate. The thin blue lines indicate the predictive distribution from 2000 posterior samples. Basis-1, Basis-2, and Basis-3 denote the mass-fractal, Guinier, and surface-scattering terms, respectively, of Equation~(\ref{eq:unified_fit}). The lower panels show the standardized residuals.}
  \label{fig:real_fitting_s2s3}
\end{figure}

Because all three samples were measured under identical USAXS conditions, we assumed that candidate noise model (e) was also appropriate for samples S2 and S3 and performed the parameter estimation accordingly.
Figure~\ref{fig:real_fitting_s2s3} shows the corresponding fits.
For each sample, the MAP curve agreed well with the data, and the standardized residuals were centered about zero with mild oscillations suggesting a small residual systematic error.
The widths of the predictive distributions are comparable to those for sample S1 in Figure~\ref{fig:real_fitting}(iii), which supports use of candidate noise model (e) for all three samples.

\subsection{Resolution of Mixing-Dependent CB Aggregate Structures}

\begin{figure}[H]
  \centering
  \includegraphics[width=\textwidth]{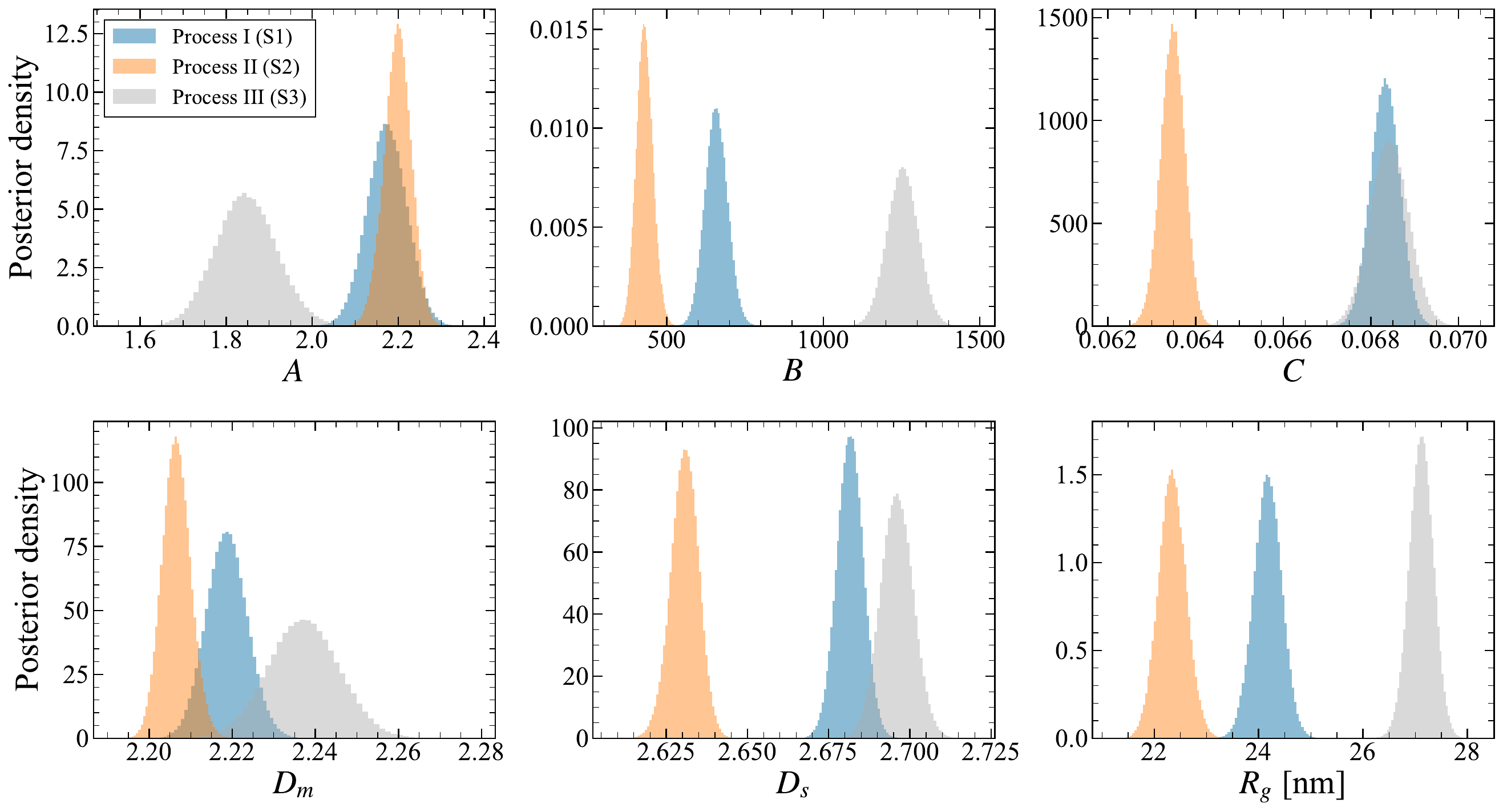}
  \caption{Posterior distributions of the structural parameters for samples S1--S3 obtained with candidate noise model (e).}
  \label{fig:real_posterior_struct}
\end{figure}

We used candidate noise model (e) to estimate the structural parameters of samples S1--S3.
Figure~\ref{fig:real_posterior_struct} compares the posterior distributions of the six structural parameters.
Although the original USAXS profiles are highly similar, the posterior distributions revealed systematic differences in the parameters describing the CB aggregate hierarchy.

For the radius of gyration $R_g$, the ordering $R_g^{(\mathrm{S2})} < R_g^{(\mathrm{S1})} < R_g^{(\mathrm{S3})}$ is clearly visible with little overlap between posteriors.
The Guinier amplitude $B$ shows an analogous ordering $B^{(\mathrm{S2})} < B^{(\mathrm{S1})} < B^{(\mathrm{S3})}$.
For the mass fractal dimension $D_m$ and surface fractal dimension $D_s$, the posteriors overlap more strongly than those of $R_g$, but the same process-dependent ordering is still observed.
Table~\ref{tab:map_ci_all} lists the MAP estimates and 95\% credible intervals of all parameters for the three samples under the selected noise model.

\input{map_ci_all_en}

The mass fractal dimension $D_m$, surface fractal dimension $D_s$, and radius of gyration $R_g$ all satisfy
\begin{equation}
    X^{(\mathrm{S2})} < X^{(\mathrm{S1})} < X^{(\mathrm{S3})} \quad (X = D_m,\ D_s,\ R_g).
    \label{eq:ordering}
\end{equation}
We used the posterior samples to compute the probability $P(X^{(\mathrm{S2})} < X^{(\mathrm{S1})} < X^{(\mathrm{S3})})$ that this ordering holds.
The results are summarized in Table~\ref{tab:ordering_prob}.

\begin{table}[H]
    \centering
    \caption{Probability that the ordering $X^{(\mathrm{S2})} < X^{(\mathrm{S1})} < X^{(\mathrm{S3})}$ holds, evaluated from the posterior samples.}
    \label{tab:ordering_prob}
    \begin{tabular}{lc}
      \toprule
      Parameter $X$ & $P(X^{(\mathrm{S2})} < X^{(\mathrm{S1})} < X^{(\mathrm{S3})})$ \\
      \midrule
      $D_m$  & $0.950$ \\
      $D_s$  & $0.989$ \\
      $R_g$  & $\approx 1.000$ \\
      \bottomrule
    \end{tabular}
\end{table}

\noindent The ordering of $D_m$ holds with the probability 95.0\%, that of $D_s$ holds with the probability 98.9\%, and that of $R_g$ holds with a probability close to unity.
Thus, samples S1--S3, which were prepared by different mixing processes, exhibited statistically credible differences in CB aggregate size, internal packing, and surface roughness.

\subsection{Comparison with Conventional Unified Fit Analysis}

Nakanishi \textit{et al.}~\cite{Nakanishi2021} fitted the same USAXS profiles using Unified Fit and reported common point estimates of $R_g = 25 \pm 2$\,nm, $D_m = 2.2 \pm 0.1$, and $D_s = 2.7 \pm 0.1$ for all three samples, where the $\pm$ values denote the fitting uncertainty from their paper.
Based on the apparent overlap of the three scattering profiles, they concluded that the dispersion of CB in the EPDM matrix is independent of the mixing process.
In contrast, the present Bayesian Unified Fit analysis yielded $R_g$ values of 22.3, 24.2, and 27.2 nm for samples S2, S1, and S3, respectively, to resolve distinct aggregate sizes among the three samples.
The differences in $D_m$ and $D_s$ were smaller in absolute magnitude but were nonetheless supported by the posterior ordering probabilities in Table~\ref{tab:ordering_prob}.
This comparison shows that the Bayesian posterior distributions provide information that is not accessible from conventional point-estimate fitting alone.
They allow small process-dependent differences to be evaluated statistically rather than judged only from the visual overlap of scattering profiles.

\subsection{Physical Interpretation of the Inferred Hierarchical Structures}

The inferred orderings can be interpreted by using the established physical meaning of each Unified Fit parameter~\cite{Beaucage1995, Beaucage1996, KohlsBeaucage2002}.
The radius of gyration $R_g$ characterizes the representative size of the scatterers and reflects the size of the CB aggregates.
The ordering $R_g^{(\mathrm{S3})} > R_g^{(\mathrm{S1})} > R_g^{(\mathrm{S2})}$ indicates that Process~III (two-roll mill, 30\,min) yielded the largest inferred aggregates, whereas Process~II (kneader, 5\,min) yielded the smallest.

The mass fractal dimension $D_m$ characterizes the density of mass distribution within the aggregate~\cite{Beaucage1996, Rieker2000}, where a larger $D_m$ corresponds to a denser internal structure.
Therefore, the ordering $D_m^{(\mathrm{S3})} > D_m^{(\mathrm{S1})} > D_m^{(\mathrm{S2})}$ suggests that Process~III formed the densest inferred aggregate structure.
The surface fractal dimension $D_s$ reflects particle-surface roughness~\cite{Beaucage1995, Rieker2000} with $D_s = 2$ corresponding to a smooth surface and $D_s \to 3$ corresponding to a rough surface.
The ordering $D_s^{(\mathrm{S3})} > D_s^{(\mathrm{S1})} > D_s^{(\mathrm{S2})}$ indicates that Process~III also exhibited the roughest aggregate surface structure.

The relative magnitudes of these variations are themselves informative.
Across the three samples, $R_g$ changed by approximately 22\% (22.31--27.16\,nm) and the Guinier amplitude $B$ changed by roughly a factor of three (425.2--1262).
In contrast, $D_m$ and $D_s$ varied by only about 1.5\% (2.206--2.239) and 2.4\% (2.631--2.695), respectively.
This pattern suggests that the mixing process predominantly affects the characteristic aggregate size and number density, while the internal mass distribution ($D_m$) and aggregate surface structure ($D_s$) are largely preserved across samples.
The near-invariance of $D_m$ and $D_s$ is consistent with the viewpoint that aggregate morphology at length scales below $R_g$ is governed primarily by CB primary-particle packing and the CB--EPDM interfacial structure rather than by the mixing energy itself.
A plausible interpretation is that Process III applies a relatively mild shear for 30\,min in the two-roll mill, which allows CB aggregates to be redistributed or reorganized in the polymer matrix without being fragmented strongly.
In contrast, Processes~I and II apply stronger shear over 5\,min and may break the aggregates into smaller fragments.
These interpretations are consistent with the inferred structural ordering, but they should be regarded as structural correlations rather than a complete mechanistic explanation.

\subsection{Consistency with Reported Tensile Properties}

Nakanishi \textit{et al.}~\cite{Nakanishi2021} reported tensile-test results for the same three samples.
Sample S3 showed the highest tensile stress at break ($\sigma_b = 18.2$\,MPa) and elongation at break ($\varepsilon_b = 344\%$) followed by sample S1 ($\sigma_b = 12.4$\,MPa, $\varepsilon_b = 300\%$) and sample S2 ($\sigma_b = 11.0$\,MPa, $\varepsilon_b = 276\%$).
Therefore, the ordering of mechanical properties matches the ordering of structural parameters (Equation~(\ref{eq:ordering})).
Because only three processes were considered, no statistical correlation analysis such as linear regression was performed. 
\begin{figure}[H]
  \centering
  \includegraphics[width=\textwidth]{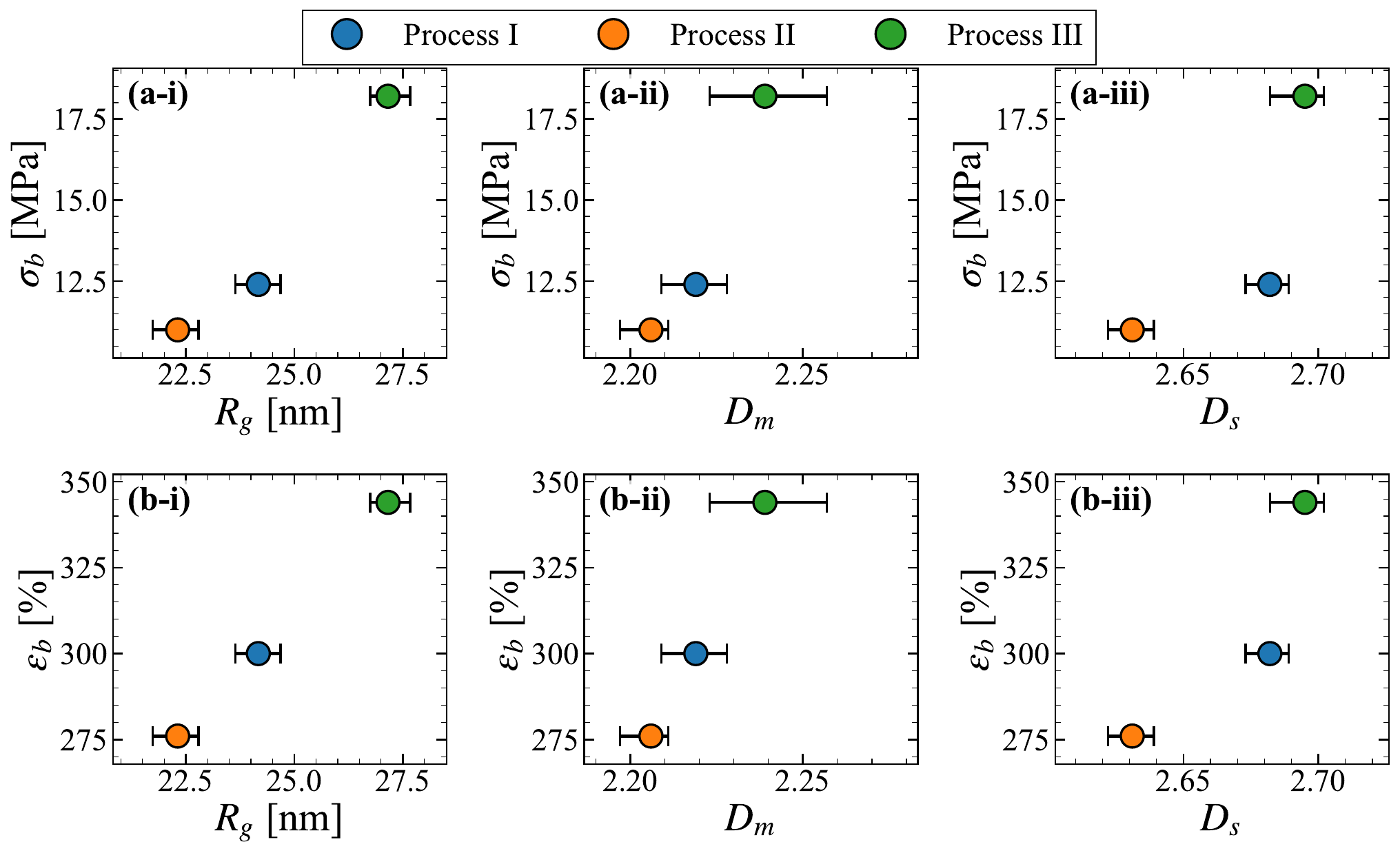}
  \caption{Relationship between CB aggregate parameters estimated by Bayesian Unified Fit ($R_g$, $D_m$, $D_s$) and reported tensile properties ($\sigma_b$, $\varepsilon_b$) of samples S1--S3 prepared by different mixing processes. Panels (a) plot $\sigma_b$ on the vertical axis, and panels (b) plot $\varepsilon_b$ on the vertical axis. The horizontal error bars represent 95\% credible intervals of the structural parameters.}
  \label{fig:structure_property_correlation}
\end{figure}
To further visualize the physical relevance of the Bayesian-inferred CB aggregate parameters, Figure~\ref{fig:structure_property_correlation} compares the MAP estimates of $R_g$, $D_m$, and $D_s$ with the reported tensile stress and elongation at break. 
Because the number of processing conditions was limited to three, we did not assign a statistical interpretation to linear regression or correlation coefficients. 
Within this limitation, both tensile properties increased monotonically with all three structural parameters, which indicates that the process-dependent differences in aggregate size, internal packing, and surface roughness resolved by Bayesian Unified Fit were consistent with the macroscopic fracture response.

Nakanishi \textit{et al.}~\cite{Nakanishi2021} attributed the differences in mechanical properties to the homogeneity and spatial cross-correlation of sulfur and zinc observed by micro X-ray fluorescence ($\mu$-XRF) and concluded that Process~III achieved the most effective vulcanization state.
The present results extend this interpretation by showing that the CB aggregate structure, which was previously treated as essentially identical among the three samples, also differed in a way that is consistent with the ordering of $\sigma_b$ and $\varepsilon_b$.
This consistency supports the physical relevance of the inferred CB aggregate parameters for rubber reinforcement.
However, we should caution that the present results indicate a correspondence between the structural and mechanical properties rather than causation.
Determining the mechanistic origin of the differences in mechanical properties will require an integrated analysis of the crosslink density, sulfur and zinc distributions, and filler-network connectivity.

Within their framework for the rational design of reinforced rubber, Kohls and Beaucage~\cite{KohlsBeaucage2002} argued that the hierarchical structural parameters of aggregates ($R_g$, $D_m$, $D_s$) serve as fundamental descriptors of the reinforcement performance.
The uncertainty-equipped $(R_g, D_m, D_s)$ values obtained here provide concrete inputs that can be used directly within this design framework.
In particular, $R_g$ showed the largest variation among the three samples and most closely tracked the ordering of tensile properties, which suggests that the aggregate size may function as a leading design parameter for controlling reinforcement through the mixing process.
This conclusion is hard to reach from point estimates alone and rests on being able to compare structural-variation magnitudes against the ordering of properties with explicit uncertainty.
More broadly, these findings indicate that the hierarchical structure of CB aggregates in EPDM rubber can retain measurable signatures of the mixing history even when the corresponding USAXS profiles appear nearly indistinguishable by conventional analysis.

\section{Conclusion}
\label{sec:conclusion}

Bayesian Unified Fit was used to quantify the effects of the mixing process on hierarchical CB structures in vulcanized EPDM rubber using previously reported USAXS profiles.
Although conventional Unified Fit treated the three samples as having essentially identical CB dispersions, posterior comparison revealed statistically credible differences in the radius of gyration, mass fractal dimension, and surface fractal dimension.
All three structural parameters followed the ordering of Process~II $<$ Process~I $<$ Process~III with high posterior probability, which indicates that the mixing process leaves measurable structural information in the aggregate size, internal packing, and surface roughness.

This structural ordering coincides with the reported ordering of the tensile stress and elongation at break, which supports the physical relevance of the inferred CB aggregate structures.
These results show that the filler morphology cannot be judged only from visually similar USAXS profiles or conventional point estimates.
The posterior uncertainty is essential for evaluating whether subtle structural differences are credible.
The present analysis employed a two-stage Bayesian workflow comprising noise-model selection followed by structural-parameter estimation, which enabled subtle inter-sample differences in structural parameters to be quantified together with their posterior uncertainty.
The present findings indicate a correspondence between the orderings of structural and mechanical properties rather than causation.
Future work integrating the crosslink density, sulfur and zinc distributions, and direct characterization of filler-network connectivity will be required to determine the mechanistic origin of the mechanical-property differences.
Nevertheless, the present findings demonstrate that uncertainty-aware Bayesian scattering analysis provides a quantitative basis for examining processing--structure--property relationships in the EPDM/CB system and potentially other polymer nanocomposites.

\section*{Acknowledgments}

This work was supported by JST BOOST, Japan Grant Number JPMJBS2418.
The authors used ChatGPT (OpenAI) to improve the English language of the manuscript.

\bibliographystyle{macromolecules_proofreading}
\bibliography{references_macromolecules_proofreading}

\clearpage
\begin{center}
  \textbf{For Table of Contents Use Only}\\[0.5em]
  \includegraphics[width=3.25in]{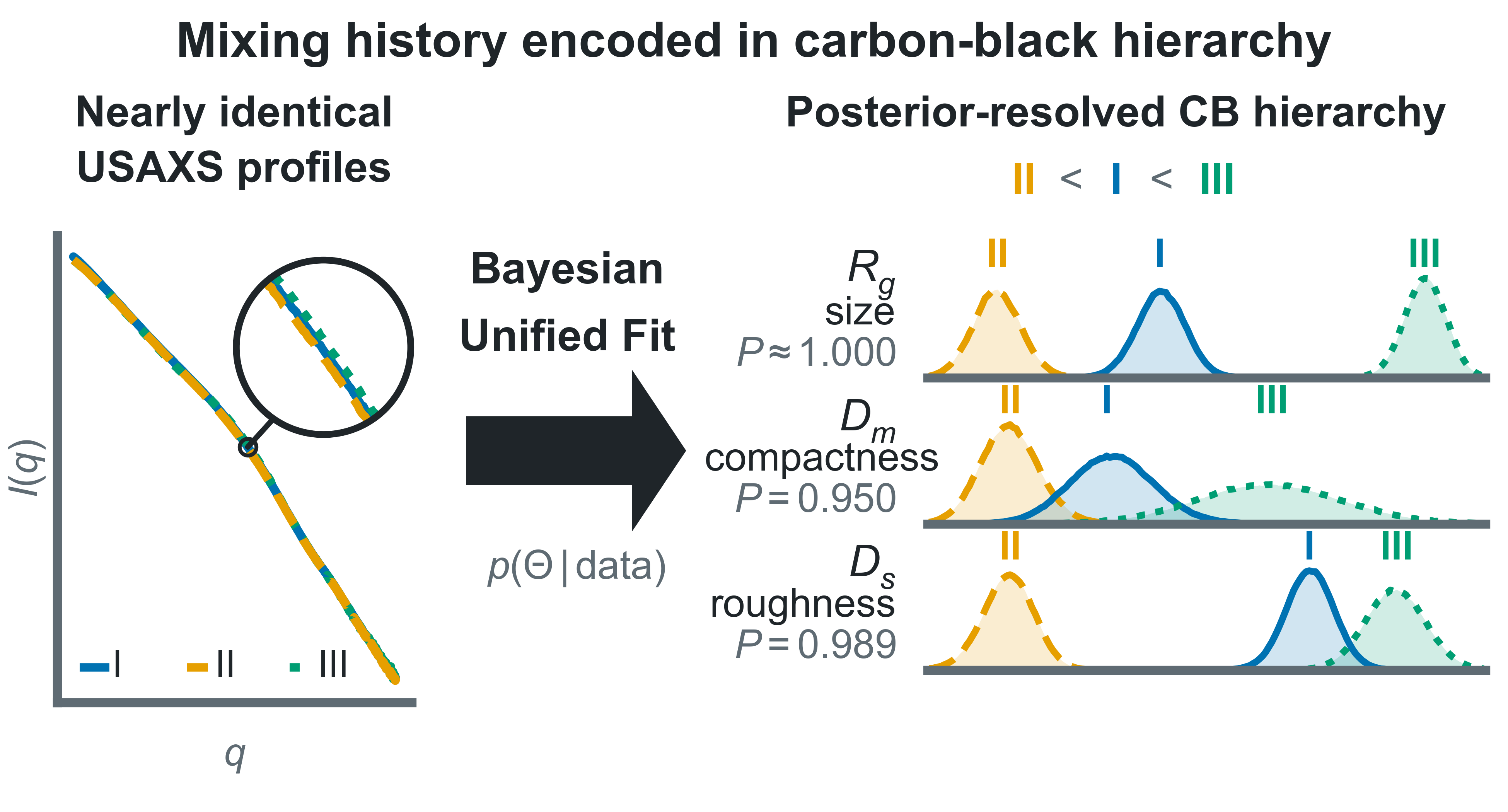}
\end{center}

\end{document}

%% file: map_ci_all_en.tex
\begin{table}[htbp]
    \centering
    \caption{MAP estimates and 95\% credible intervals for samples S1--S3.}
    \label{tab:map_ci_all}
    \footnotesize
    \setlength{\tabcolsep}{4pt}
    \begin{tabular}{lccc}
      \toprule
      Param. & Process I (S1) & Process II (S2) & Process III (S3) \\
      \midrule
      $A$ & $2.174\ [2.084,\ 2.265]$ & $2.208\ [2.157,\ 2.282]$ & $1.834\ [1.678,\ 1.959]$ \\
      $B$ & $657.0\ [584.9,\ 725.4]$ & $425.2\ [365.1,\ 471.1]$ & $1262\ [1172,\ 1371]$ \\
      $C$ & $0.0683\ [0.0675,\ 0.0689]$ & $0.0635\ [0.0629,\ 0.0640]$ & $0.0683\ [0.0672,\ 0.0690]$ \\
      $D_m$ & $2.219\ [2.209,\ 2.228]$ & $2.206\ [2.197,\ 2.211]$ & $2.239\ [2.223,\ 2.257]$ \\
      $D_s$ & $2.682\ [2.673,\ 2.689]$ & $2.631\ [2.622,\ 2.639]$ & $2.695\ [2.682,\ 2.702]$ \\
      $R_g$ [nm] & $24.17\ [23.64,\ 24.68]$ & $22.31\ [21.73,\ 22.79]$ & $27.16\ [26.75,\ 27.67]$ \\
      \midrule
      $\alpha_1$ [$\times 10^{-4}$] & $2.99\ [2.31,\ 3.98]$ & $1.33\ [1.00,\ 1.74]$ & $6.70\ [5.11,\ 8.42]$ \\
      $\alpha_2$ [$\times 10^{-4}$] & $3.93\ [3.51,\ 4.57]$ & $2.98\ [2.68,\ 3.42]$ & $6.75\ [6.02,\ 7.84]$ \\
      \bottomrule
    \end{tabular}
    \par\smallskip
    \begin{minipage}{\textwidth}
      \footnotesize\raggedright
      Notes: Values in brackets denote the lower and upper bounds of the 95\% credible interval.
    \end{minipage}
\end{table}